\documentclass[sigconf]{acmart}
\usepackage{booktabs} 
\usepackage{amsmath, amsthm, amssymb}
\usepackage{comment}
\usepackage{multirow}

\usepackage{rotating}
\usepackage{booktabs}
\usepackage{tabularx}
\usepackage{subfigure}

\usepackage{tikz}
\usepackage{pgfplots}
\usetikzlibrary{patterns}
\usetikzlibrary{matrix}
\usepackage{enumitem}

\newcommand{\nb}[3]{
  \fcolorbox{black}{#2}{\bfseries\sffamily\scriptsize#1}
    {\sf\small$\blacktriangleright$\textit{#3}$\blacktriangleleft$}
}

\newcommand\jie[1]{\nb{Jie}{red}{#1}}
\newcommand\wenjie[1]{\nb{Wenjie}{green}{#1}}

 \acmConference[CIKM'17]{International Conference
 on Information and Knowledge Management}{November 2017}{Pan Pacific Singapore} 

\begin{document} 


\fancyhead{}
\settopmatter{printacmref=false, printfolios=false}

\ccsdesc[500]{Information systems~Information retrieval}
\ccsdesc[500]{Information systems~Recommender systems}
\ccsdesc[500]{Computing methodologies~Neural networks}

\title{Interacting Attention-gated Recurrent Networks for Recommendation}

\renewcommand{\shorttitle}{Interacting Attention-gated Recurrent Network}


\author{Wenjie Pei}
\authornote{Both authors contributed equally to the paper.}
\affiliation{%
  \institution{Delft University of Technology}
  \city{Delft} 
  \country{The Netherlands} 
}
\email{w.pei-1@tudelft.nl}

\author{Jie Yang}
\authornotemark[1]
\affiliation{%
  \institution{Delft University of Technology}
  \city{Delft} 
  \country{The Netherlands} 
}
\email{j.yang-3@tudelft.nl}

\author{Zhu Sun}
\affiliation{%
  \institution{Nanyang Technological University}
  \country{Singapore}}
\email{sunzhu@ntu.edu.sg}

\author{Jie Zhang}
\affiliation{%
  \institution{Nanyang Technological University}
  \country{Singapore}}
\email{zhangj@ntu.edu.sg}

\author{Alessandro Bozzon}
\affiliation{%
  \institution{Delft University of Technology}
  \city{Delft} 
  \country{The Netherlands} 
}
\email{a.bozzon@tudelft.nl}

\author{David M.J. Tax}
\affiliation{%
  \institution{Delft University of Technology}
  \city{Delft} 
  \country{The Netherlands} 
}
\email{d.m.j.tax@tudelft.nl}


\begin{abstract}
Capturing the temporal dynamics of user preferences over items is important for recommendation. Existing methods mainly assume that all time steps in user-item interaction history are equally relevant to recommendation, which however does not apply in real-world scenarios where user-item interactions can often happen accidentally. More importantly, they learn user and item dynamics separately, thus failing to capture their joint effects on user-item interactions. To better model user and item dynamics, we present the Interacting Attention-gated Recurrent Network (IARN) which adopts the attention model to measure the relevance of each time step. In particular, we propose a novel attention scheme to learn the attention scores of user and item history in an interacting way, thus to account for the dependencies between user and item dynamics in shaping user-item interactions. By doing so, IARN can selectively memorize different time steps of a user's history when predicting her preferences over different items. Our model can therefore provide meaningful interpretations for recommendation results, which could be further enhanced by auxiliary features. Extensive validation on real-world datasets shows that IARN consistently outperforms state-of-the-art methods.

\end{abstract}

%
%

\ccsdesc[500]{Information systems~Information retrieval}
\ccsdesc[500]{Information systems~Recommender systems}
\ccsdesc[500]{Computing methodologies~Neural networks}


\keywords{Recurrent Neural Network, Attention Model, User-item Interaction, Feature-based Recommendation}

\maketitle

\section{Introduction}

Recommendation is a fundamental task to enable personalized information filtering, thus to mitigate the information overload problem~\cite{su2009survey}. The goal is to learn user preferences from historical user-item interactions, based on which recommend relevant items. In reality, user preferences often evolve over time, affected by dynamic user inclinations, item perception and popularity. Temporal context therefore has been recognized as an important type of information for modeling the dynamics of user preferences. It has extensive applications, ranging from movie recommendation \cite{bennett2007netflix}, music recommendation \cite{koenigstein2011yahoo}, to location recommendation \cite{yuan2013time}.  

Most existing methods \cite{koren2009collaborative,koenigstein2011yahoo, pham2015general,yuan2013time,gao2013exploring} model the temporal dynamics by extending the latent factor model (LFM) \cite{shi2014collaborative} with handcrafted features, so as to describe certain temporal patterns of user-item interactions. For example, they either bin user-item interactions into time windows, assuming similar user behavioral patterns in the same window \cite{koenigstein2011yahoo,yuan2013time}, or adopt a time decay function to under-weight the interactions occurring deeper into the past \cite{koren2009collaborative,pham2015general}. The handcrafted features, though proven to be effective, cannot capture complex temporal patterns in reality  \cite{wu10recurrent}. More importantly, these methods cannot automatically select important interaction records in user-item interaction history when modeling user preferences. This greatly limits their application in real-world scenarios where user-item interactions can often happen accidentally. 


Recently, recurrent neural network (RNN) \cite{rumelhart1988learning} based methods have emerged as a promising approach to model the temporal dynamics of user preferences \cite{hidasi2015session,jing2017neural,wu10recurrent}. RNN captures both the latent structures in historical user-item interactions -- through hidden units -- and their dynamics along the temporal domain. Unlike LFM based methods, these methods are nonparametric, thus can learn inherent dynamics that are more complex and suitable for making recommendations.  A specific type of \emph{gated} RNN, i.e. Long Short-Term Memory (LSTM) \cite{hochreiter1997long}, is employed by the state-of-the-art recommendation method \cite{wu10recurrent} to model both user and item dynamics. The gating mechanism is adopted to balance the information flow from the current and previous time steps, thus can more effectively preserve historical information over time for recommendation. 

\begin{figure}
	\includegraphics[width=0.49\textwidth]{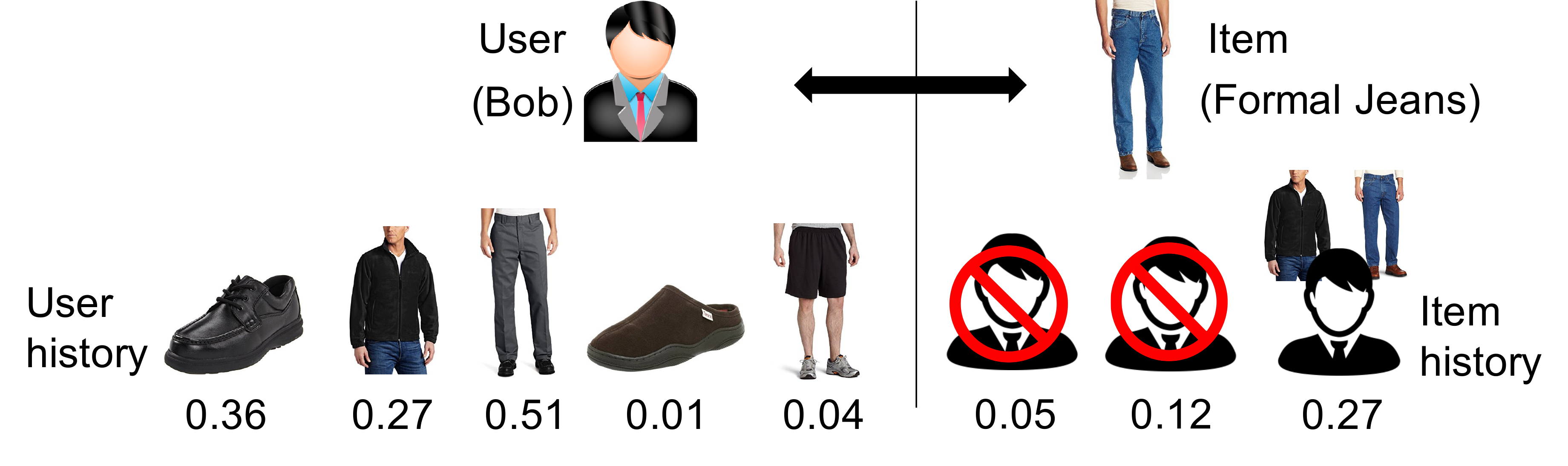}
    \caption{Example of user-item interactions determined by dependent user dynamics and item dynamics. 
    The numbers below user and item history are the attention scores inferred by our proposed model, which are used to select relevant time steps in user and item history to accurately predict the user's preference over the item.}
    \label{fig:illustration}
    \vspace{-0.15in}
\end{figure}

Nevertheless, LSTM models the gate w.r.t. each hidden unit instead of the whole time step, making it difficult to interpret the importance of each time step for the final recommendation. More importantly, gates for modeling user dynamics and item dynamics so far are learned separately. In real-world scenarios, however, user and item dynamics are dependent on each other, and can jointly affect user-item interactions.  
Consider the target user Bob in Figure~\ref{fig:illustration}, who is interested in both formal clothing (e.g., leather shoes and trousers) and casual clothing (e.g., casual shoes and shorts), as described by his purchasing history. We observe that Bob buys a pair of formal jeans, which were historically bought by users with various interests. The interaction between Bob and the formal jeans is therefore determined by his interest in formal clothing and the inherent property of the formal jeans, namely, the formal style. Such an interest and property could only be learned from historical user-item interactions when no additional auxiliary features are given. Therefore, to accurately capture Bob's preference over the formal jeans, the recommendation model should be able to identify the important time steps of Bob's purchasing history when he bought formal clothing. Similarly, in the history of formal jeans it should be able to identify time steps when they were bought by users who are also interested in formal style clothing, thus to capture the item property relevant to Bob's interest. 

In this paper, we introduce the Interacting Attention-gated Recurrent Network (IARN) which adopts the attention model to measure the relevance of each time step of user history and item history for recommendation. In particular, we propose a novel attention scheme which allows IARN to learn the relevance -- measured by attention scores -- of time steps in user and item history in an interacting way, so as to capture the dependencies between user and item dynamics in shaping user-item interactions. As a result, IARN can selectively memorize different time steps of a user's history when predicting her preferences over different items, thereby providing meaningful interpretations for the prediction. For instance, attention scores learned by IARN for the example in Figure~\ref{fig:illustration} are shown under the user and item history in the figure (note this example is based on our results on a real-world dataset). 

IARN could be further enhanced by incorporating auxiliary features of users or items. In this paper we provide methods to integrate IARN with auxiliary features organized in a flat or a hierarchical structure. 
More specifically, our main contributions include: 
\begin{itemize}[noitemsep,nolistsep,leftmargin=*]
\item We extend recurrent networks for modeling user and item dynamics with a novel gating mechanism, which adopts the attention model to measure the relevance of individual time steps of user and item history for recommendation. 
\item We design a novel attention scheme which allows the user- and item-side recurrent networks to interact with each other, thus to capture the dependencies between user and item dynamics to improve recommendation accuracy and interpretability. 
\item We propose the IARN method implementing the interacting attention-gate as described above, and show how it can be further enhanced by auxiliary features organized in different structures.
\item We conduct extensive experiments to evaluate the proposed IARN method on six real-world datasets, demonstrating that IARN consistently outperforms state-of-the-art methods. 
\end{itemize}

\section{Related Work}

This section provides an overview of state-of-the-art recommendation methods related to our work. We review them from two orthogonal perspectives: 
(1) the underlying recommendation models; (2) the incorporation of side information  for recommendation. 

\subsection{Underlying Recommendation Models}

The recently pervasive recommendation methods can be broadly categorized into two types, namely, the latent factor model based methods and the neural network based ones.

\smallskip\noindent\textbf{Latent Factor Model.}
Due to the high efficiency, state-of-the-art recommendation methods have been dominated by Latent Factor Model (LFM). It decomposes the high-dimensional user-item rating matrix into low-dimensional user and item latent matrices. A panoply of algorithms have been proposed to date based on LFM, including matrix factorization (MF) \cite{koren2009matrix},
Bayesian personalized ranking (BPR) \cite{rendle2009bpr}, collective matrix factorization (CMF) \cite{singh2008relational},  factorization machine (FM) \cite{rendle2010factorization}, SVD++ \cite{koren2008factorization}, to name a few. 
Despite of their success, LFM based methods suffer from the following essential limitations. First of all, they merely leverage global statistical information of user-item interaction data, while cannot capture fine-grained regularities in the latent factors \cite{pennington2014glove}. Second, LFM based recommendation methods generally learn latent representations of users and items in a linear fashion, which may not be always suitable in real-world scenarios. 
Besides, most LFM based methods ignore the temporal dynamics of user preferences, assuming that the future user-item interactions are known in advance, which is contradictory with the real-world application. There are a few LFM based methods specifically designed for fusing temporal information, which will be reviewed in section 2.2. 

\smallskip\noindent\textbf{Neural Networks.}
Stemming from the success in related domains (e.g., computer vision, speech recognition, and natural language processing), Neural Network (NN) based methods have recently attracted a considerable amount of interests from the recommendation community. In contrast to LFM based recommendation methods, NN based methods have shown to be highly effective in capturing local item relationships by modeling item co-occurrence in individual users' interaction records. Typical methods are User2Vec \cite{grbovic2015commerce} and Item2Vec \cite{barkan2016item2vec}, which are inspired by word embedding techniques \cite{mikolov2013efficient,mikolov2013distributed}.
Furthermore, NN based models can learn nonlinear latent representations through the activation functions (e.g., sigmoid, ReLU \cite{nair2010rectified}).
For instance, Suvash et al. propose the AutoRec \cite{sedhain2015autorec} recommendation method based on autoencoders \cite{hinton2006reducing}. He et al. propose neural collaborative filtering \cite{heneural} to learn non-linear interactions between users and items. 
Recently, the Recurrent Neural Network (RNN) based methods \cite{wu10recurrent, jing2017neural, hosseini2017recurrent, hidasi2015session} have gained significant enhancement in recommendation thanks to the ability of preserving historical information over time for recommendation.  These methods learn time-varying representations of users/items (i.e., hidden-states) in each time step, by taking into account both the present and historical data. 
The learned states can be used for generating recommendations for the future, therefore being more realistic and attractive for real-world applications.
To sum up, NN based methods possess essential advantages and have shown to be more effective to enhance recommendation performance.  

\subsection{Incorporating Side Information}

To better model user preferences thus to further improve recommendation performance, many researchers endeavor to incorporate side information, i.e., information complementing user-item interactions, into recommendation models. Here we focus on the literature with consideration of two types of side information related to our work, i.e., temporal context and auxiliary features.

\smallskip\noindent\textbf{Temporal Context.}
It has been well recognized that user preferences change over time. This can be due to drifting user inclinations for item, or the constantly changing item perception and popularity when new selection emerges \cite{koren2009collaborative,jing2017neural}. Hence, recommendation methods that capture temporal dynamics of user preferences could provide improved recommendation performance. In the branch of LFM based methods, some take temporal information into consideration based on time windows, assuming user-item interactions in the same window have similar patterns. For instance, Koenigstein et al. \cite{koenigstein2011yahoo} and Yuan et al. \cite{yuan2013time} propose such methods for music and Point-of-Interest recommendation. A disadvantage is that they regard all interactions within the considered time window equally,  completely ignoring the relationships of interactions among different windows. In addition, binning user-item interactions aggravates the data sparsity problem. Some other LFM based methods attempt to address these issues by adopting a time decay function to under-weight the instances as they occur deeper into the past. These include TimeSVD++ proposed by Koren \cite{koren2009collaborative} and HeteRS proposed by Pham et al. \cite{pham2015general}. However, these methods could not capture other types of temporal patterns, e.g., certain user-item interactions could be driven by the long-term interest of a user which could not be modeled in a decay manner. In fact, all LFM based methods handle temporal context by creating handcrafted features, thus cannot capture complex temporal patterns in reality. 

Contrarily, RNN based methods are nonparametric, thus can learn inherent dynamics of user preferences that are more complex. For instance, Hidasi et al. \cite{hidasi2015session} propose a RNN based approach for session-based recommendation. Hosseini et al. \cite{hosseini2017recurrent} introduce a recurrent Poisson factorization framework for recommendation.  Among different RNN models, Long Short-Term Memory (LSTM) \cite{hochreiter1997long} has gained much popularity in recommendation due to their capability in dealing with the gradient vanishing problem \cite{zaremba2014recurrent}. Jing et al. \cite{jing2017neural}  present a LSTM based method  to estimate when a user will return to a site and what her future listening behavior will be.
Wu et al. \cite{wu10recurrent} propose a LSTM based method, i.e., recurrent recommender network (RRN), to model user and item dynamics. This is the most closely related work to ours. However, one major shortcoming of these gated RNN based methods is that the learned gate lacks interpretability, limiting further improvements of recommendation accuracy.  More importantly, these methods model user and item dynamics separately, thus failing to capture their dependencies and their joint effects on user-item interactions. 

Motivated by the attention scheme in human foveal vision, attention mechanism has been employed by NN based methods to cope with the data noisy problem by identifying relevant parts of the input for the prediction task. It has been applied in a broad spectrum of disciplines, from natural language processing \cite{yang2016hierarchical} to computer vision \cite{pei2017temporal, xu2015show}. However, how to effectively exploit the attention mechanism in recommender systems is still an open research question. To the best of our knowledge, we are the first to propose recurrent network based recommendation method that integrates attention mechanism to automatically learn the relevance of individual time steps for recommendation, so as to enhance both recommendation interpretability and accuracy. More importantly, we design a novel attention scheme that allows user- and item-side recurrent networks to interact with each other, thus to capture the dependencies between user and item dynamics and their joint effects on user-item interactions.

\smallskip\noindent\textbf{Auxiliary Features.}
Better representations of users and items can also be obtained by incorporating auxiliary features into recommendation, due to the rich semantic information encoded by them. 
Most existing feature-based recommendation approaches are built upon LFM. These methods are either designed to incorporate features in a flat structure or a hierarchy. For instance, the popular CMF \cite{singh2008relational} and FM \cite{rendle2010factorization} are designed for integrating flat features into recommendation. Recently it has been found that feature hierarchies, i.e., hierarchically organized features, can be more effective in boosting the accuracy as well as the interpretability of recommendation. 
He et al. \cite{he2016sherlock} devise a visually-aware recommendation model by manually defining the feature hierarchy influence on items. Yang et al. \cite{yang2016learning} design a recommendation method that automatically learns feature hierarchy influence on user/item by a parameterized regularization traversing from root to leaf features. More recently, Sun et al. \cite{sun2017exploiting} introduce  a unified recommendation framework that seamlessly incorporates both vertical and horizontal dimensions of feature hierarchies for effective recommendation. 
In this paper, we show how to incorporate features organized in both flat and hierarchical structures into our model. 
Note that although in other domains like nature language processing, a few work \cite{yang2016hierarchical} attempts to integrate hierarchies into RNN model, there is few such kind of approach in recommendation. Hence, we are the first to explore the effect of features organized in different structures together with recurrent networks to learn optimal representations of users and items for improved recommendation interpretability and accuracy.

\section{Interacting Attention-gated Recurrent Networks}

\begin{figure*}
	\includegraphics[width=0.95\textwidth]{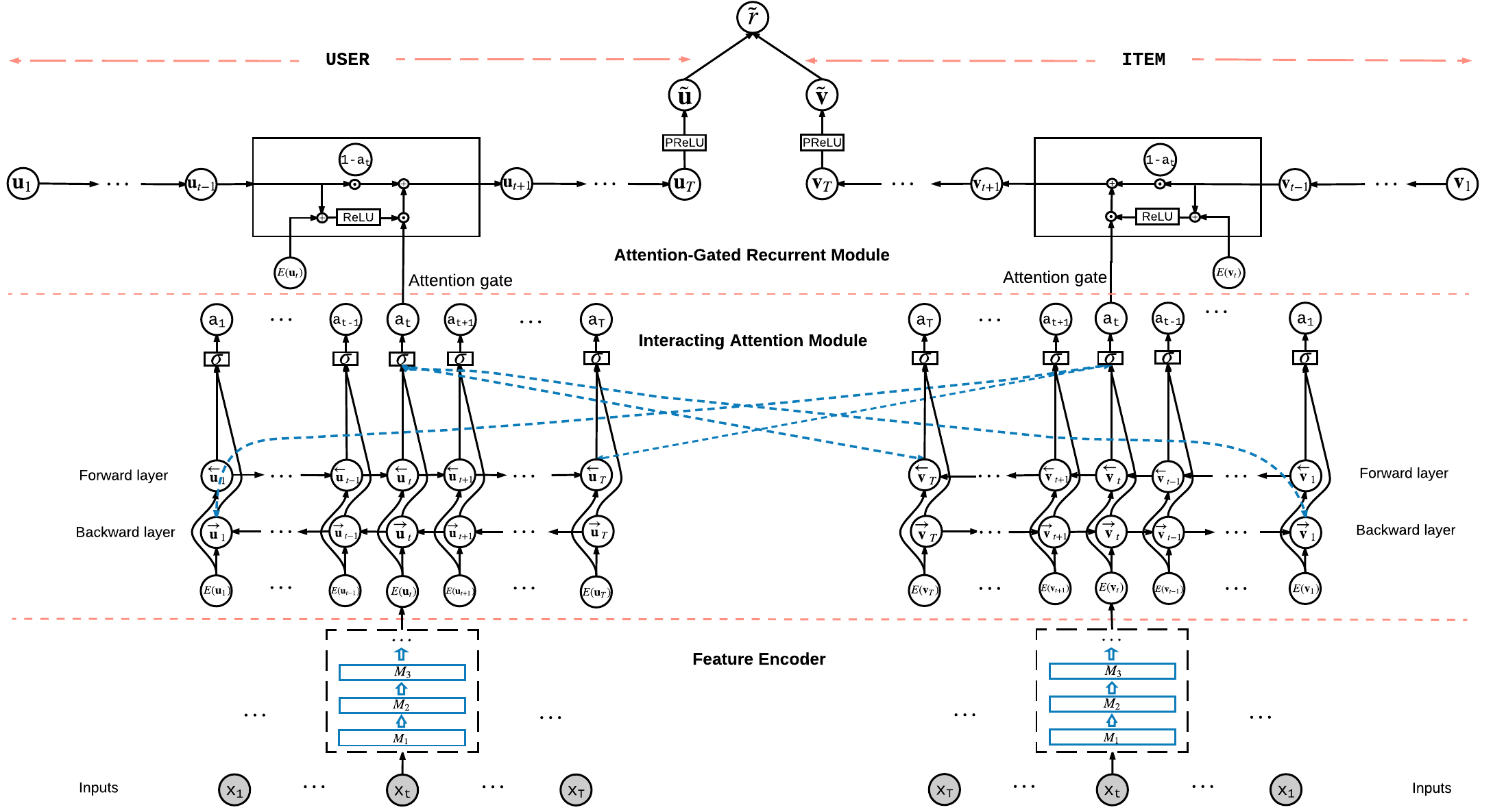}
    \setlength{\abovecaptionskip}{0.5pt}
    \caption{Architecture of IARN. Two recurrent networks are employed to learn hidden representations of users and items. Each recurrent network is composed of an Attention-gated Recurrent Module to capture the user/item dynamics, an Interacting Attention Module that models the attention, and a Feature Encoder to extract semantically rich information from the input data. For the sake of clarity, here we only show the connections from a single attention gate in the user (item) network to the forward and backward layers of the item (user) network, however the attention gates of all time steps in the user (item) network are connected to forward and backward layers of the item (user) network.  Overall, the two recurrent networks interact with each other to learn the dependencies between user and item dynamics. }
    \label{fig:architecture}
    \vspace{-0.05in}
\end{figure*}

Given the historical user-item interaction data as the input, we aim to learn high-quality hidden representations for both users and items, which are then used for subsequent recommendation. The extracted representations are expected to: 1) capture the temporal dynamics contained in both user and item history with physical interpretability of each time step; 2) learn the dependencies between user and item dynamics in shaping user-item interactions; 3) extract semantically rich information from training data through the incorporation of auxiliary features. To achieve these goals, we propose the Interacting Attention-gated Recurrent Network (IARN), which is composed of three modules: Attention-gated Recurrent Module, Interacting Attention Module and Feature Encoder. They are designed correspondingly to the three aforementioned goals.

The overall architecture of IARN is illustrated in Figure~\ref{fig:architecture}. It employs two recurrent networks to learn  compact and effective hidden representations for the paired user and item, respectively. Each recurrent network is composed of an Attention-gated Recurrent Module, an Interacting Attention Module, and a Feature Encoder. Instead of behaving independently, the two recurrent networks interact with each other to model the dependencies between user and item  dynamics.

\smallskip
\noindent\textbf{Input and Output Layers.} We first describe the input and output layer of IARN, then in the following subsections we will elaborate on the three modules in a top-down fashion to explain step by step how the input and output layers are connected to achieve our goal. 

Let $\mathcal{U}$ and  $\mathcal{V}$ be the user and item set, respectively. As input, each user $i\in \mathcal{U}$ is described by a sequence $\mathbf{x}_i$, which contains the representations (e.g., the embeddings) of all the items rated by her, ordered by the rating time. Similarly, each item $j\in \mathcal{V}$ is described by a sequence $\mathbf{x}_j$ that contains the representations of all users who have rated the item, ordered by the rating time. Through the three modules, IARN learns from $\mathbf{x}_i$ and $\mathbf{x}_j$ the \emph{hidden representation} of user $i$, denoted by $\tilde{\mathbf{u}}_{i}$, and the \emph{hidden representation} of item $j$, denoted by  $\tilde{\mathbf{v}}_{j}$. $\tilde{\mathbf{u}}_{i}$ and  $\tilde{\mathbf{v}}_{j}$ are then used to predicted user $i$'s preference rating $\tilde{r}_{ij}$ over item $j$ via an inner product operation:
%
%
\begin{equation}
\vspace{-0.05in}
\tilde{r}_{ij} = \langle \tilde{\mathbf{u}}_{i}, \tilde{\mathbf{v}}_{j} \rangle
\end{equation}

\subsection{Attention-gated Recurrent Module}
In order to learn high-quality hidden representations $\tilde{\mathbf{u}}_{it}$ and  $\tilde{\mathbf{v}}_{jt}$, we propose the Attention-gated Recurrent Module to preserve the information of previous time steps with relevance modeled by attention scores, which are obtained by the Interactive Attention Module that will be present in section~\ref{sec:IAM}. 
Specifically, we construct two attention-gated recurrent modules for the paired user and item, respectively. It should be noted that these two modules do not share  parameters, since users and items are not expected to share similar hidden representations. This makes our method different from Siamese Network~\cite{bromley1993signature},  a well-known method for object  comparison.

Both user- and item-side Attention-gated Recurrent Modules contain two layers, namely, a recurrent layer and a fully-connected layer. The recurrent layer models the temporal dynamics of users and items as \emph{hidden-states}, while the fully-connected layer transform the hidden-states of users and items in the last time step to the \emph{hidden representations} for prediction. We first describe the full-connected layer, then introduce in detail the recurrent layer. 

\smallskip\noindent\textbf{User- and Item-Side Fully-Connected Layers.} Denote the last hidden-states of user- and item-side recurrent layers as $\mathbf{u}_{iT_i}$ and $\mathbf{v}_{jT_j}$, respectively. The hidden representations $\tilde{\mathbf{u}}_{i}$ and $\tilde{\mathbf{v}}_{j}$ are transformed from these hidden-states by non-linear transformations:
\begin{equation}
\vspace{-0.05in}
\begin{cases}
\tilde{\mathbf{u}}_{i} = g(\widetilde{\mathbf{W}}_u \cdot \mathbf{u}_{iT_i} + \tilde{b}_u) \\
\tilde{\mathbf{v}}_{j} = g(\widetilde{\mathbf{W}}_v \cdot \mathbf{v}_{jT_j} + \tilde{b}_v) 
\end{cases}
\end{equation}
\noindent Herein, $\widetilde{\mathbf{W}}_u$ and $\widetilde{\mathbf{W}}_v$ are linear transformation parameters of the user- and item-side layers, respectively; $\tilde{b}_u$ and $\tilde{b}_v$ are the bias terms; $g$ is the activation function, for which we use the Parametric Rectified Linear Unit (PReLU) \cite{he2015delving}. PReLU allows the output of the unit to be either positive or negative, thus is more suitable for representing users/items  --  intuitively, a user could either like or dislike certain types of items (e.g., action movies), and an item could either be of a specific type or not.

\smallskip
\noindent\textbf{User-Side Attention-gated Recurrent Layer.} 
Given the user $i$ whose corresponding input sequence is $\mathbf{x}_i = \{\mathbf{x}_{i1}, \mathbf{x}_{i2}, \ldots\}$. We denoted the attention score at time step $t$ by $a_{it}$, which is a scalar value between $[0,1]$ inferred by the Interacting Attention Module. The hidden-state of user $i$ at time $t$ is then modeled as
\begin{equation}
\vspace{-0.05in}
\mathbf{u}_{it} = (1-a_{it}) \cdot \mathbf{u}_{i(t-1)} + a_{it} \cdot \mathbf{u}_{it}'
\label{equ:gate}
\end{equation}
\noindent where $\mathbf{u}_{i(t-1)}$ is the hidden-state in the previous time step and $\mathbf{u}_{it}'$ is the candidate state value obtained by fully incorporating the input at the current time step:
\begin{equation}
\mathbf{u}_{it}' = g(\mathbf{W}_u \cdot \mathbf{u}_{i(t-1)} + \mathbf{H}_u \cdot E_u(\mathbf{x}_{it}) + b_u)
\end{equation}
\noindent where $\mathbf{W}_u$ and $\mathbf{H}_u$ are respectively the linear transformation parameters for the previous and current time steps; $b_u$ is the bias term; and $E_u(\cdot)$ is the Feature Encoder that transforms the original user sequence by considering auxiliary features, which will be detailed in section~\ref{sec:HFE}. We use ReLU for the activation function $g$.

Equation~\ref{equ:gate} balances the contributions of the input of the current candidate hidden-state and the previous hidden-state with an attention gate described by the attention score $a_{it}$. Attention gates with high scores will focus more on the current input than previous hidden-states, while recurrent gates with low attention scores will ignore the current input and inherit more information from previous time steps. The attention score therefore quantifies the importance of individual time steps in the final prediction. 

\smallskip
\noindent\textbf{Item-Side Attention-gated Recurrent Layer.} Similarly, for the item-side recurrent layer, we model the the hidden-state as follows 
\begin{equation}
\begin{cases}
\mathbf{v}_{jt} = (1-a_{jt}) \cdot \mathbf{v}_{j(t-1)} + a_{jt} \cdot \mathbf{v}_{jt}' \\
\mathbf{v}_{jt}' = g(\mathbf{W}_v \cdot \mathbf{v}_{j(t-1)} + \mathbf{H}_v \cdot E_v(\mathbf{x}_{jt}) + b_v)
\end{cases}
\end{equation}
\noindent where $\mathbf{x}_{jt}$ is the input of item $j$ at time $t$; $\mathbf{W}_v$, $\mathbf{H}_v$, and $b_v$ are the network parameters; $a_{jt}$ is the attention score that serves as attention gate; and $E_v(\cdot)$ is the Feature Encoder for transforming item sequences, introduced in section~\ref{sec:HFE}.

\subsection{Interacting Attention Module}
\label{sec:IAM}
We propose the Interacting Attention Module for both users and items to measure the saliency and relevance of the input in each time step to rating prediction. The key point in this module is that the inferred attention score should not only consider the current time step in the sequence on its own side, but also take into account the information of the other side so as to model the interacting dependency between the paired user and item. 

\smallskip\noindent\textbf{User-Side Interacting Attention Module.} To maximize the utility of the input sequence, we model the saliency score based on both the input observation at the current time step and the information from neighboring observations in both directions. This is achieved by using a bi-directional RNN \cite{schuster1997bidirectional}, which includes a forward layer and a backward layer, as depicted in Figure~\ref{fig:architecture}. The attention score $a_{it}$ at time step $t$ in Equation~\ref{equ:gate} on the user side is  modeled as: 
\begin{equation}
a_{it} = \sigma (\mathbf{M}_u{^\top} \cdot \tanh(\mathbf{L}_u \cdot (\overset\rightarrow{\mathbf{u}}_{it}; \overset\leftarrow{\mathbf{u}}_{it}; \overset\rightarrow{\mathbf{v}}_{jT_j};  \overset\leftarrow{\mathbf{v}}_{j1}) + b_u' )))
\label{equ:iam}
\end{equation}
Wherein a two-layer network is used to calculate the attention score: $\mathbf{L}_u$ is a matrix as the parameter of the fusion layer that fuses both directional layers of our bi-directional RNN; $b_u'$ is the bias term of the fusion layer; and  $\mathbf{M}_u$ is the weight vector of the second layer; $\sigma$ is sigmoid function applied as the activation function to control the attention score to lie between $[0,1]$; $(;)$ denotes the concatenation among vectors;  $\overset\rightarrow{\mathbf{u}}_{it}$ and  $\overset\leftarrow{\mathbf{u}}_{it}$ perform as the summary of context information around time step $t$ in the user sequence $x_i$. Specifically, 
\begin{equation}
\vspace{-0.05in}
\begin{split}
\overset\rightarrow{\mathbf{u}}_{it} = g(\overrightarrow{\mathbf{W}}_u\cdot E_u(\mathbf{x}_{it}) + \overrightarrow{\mathbf{H}}_u\cdot \overset\rightarrow{\mathbf{u}}_{i(t-1)} + \overset\rightarrow{b}_u)\\
\overset\leftarrow{\mathbf{u}}_{it} = g(\overleftarrow{\mathbf{W}}_u\cdot E_u(\mathbf{x}_{it}) + \overleftarrow{\mathbf{H}}_u\cdot \overset\leftarrow{\mathbf{u}}_{i(t+1)} + \overset\leftarrow{b}_u)
\end{split}
\end{equation}
\noindent Therefore, $\overset\rightarrow{\mathbf{u}}_{it}$ summarizes the sequence from the beginning to time $t$, while $\overset\leftarrow{\mathbf{u}}_{it}$ summarizes the sequence from the end to time $t$. 

Similarly, $\overset\rightarrow{\mathbf{v}}_{jT_j}$
$\overset\leftarrow{\mathbf{v}}_{j1}$ in Equation~\ref{equ:iam} are the summary of the paired item sequence $\mathbf{x}_j$, whose calculation will be introduced later in detail by Equation~\ref{equ:bi2}. They are concatenated together with the summary of the user-side sequence, and used as input of the fusion layer. In this way, the resulting attention score $a_{it}$ is used to characterize the relevance of the current time step $t$ of user sequence $\mathbf{x}_i$ conditioned on the paired item sequence $\mathbf{x}_j$. 

\smallskip\noindent\textbf{Item-Side Interactive Attention Module.} Similarly, for item-side, we have 
\begin{equation}
a_{jt} = \sigma (\mathbf{M}_v{^\top} \cdot \tanh(\mathbf{L}_v \cdot (\overset\rightarrow{\mathbf{v}}_{jt}; \overset\leftarrow{\mathbf{v}}_{jt}; \overset\rightarrow{\mathbf{u}}_{iT_i};  \overset\leftarrow{\mathbf{u}}_{i1}) + b_v')))
\label{eqn:att_v}
\end{equation}
where $\mathbf{L}_v, b_v'$ are the parameters of the fusion layer, and
$\mathbf{M}_v$ is the weight vector of the second layer; $\overset\rightarrow{\mathbf{v}}_{jt}$ and $\overset\leftarrow{\mathbf{v}}_{jt}$ perform as the summary of the context information around time step $t$ in the item sequence $\mathbf{x}_j$:
\begin{equation}
\begin{split}
\overset\rightarrow{v}_{jt} = g(\overrightarrow{\mathbf{W}}_v\cdot E_v(\mathbf{x}_{jt}) + \overrightarrow{\mathbf{H}}_v\cdot \overset\rightarrow{\mathbf{v}}_{j(t-1)} + \overset\rightarrow{b}_v)\\
\overset\leftarrow{\mathbf{v}}_{jt} = g(\overleftarrow{\mathbf{W}}_v\cdot E_v(\mathbf{x}_{jt}) + \overleftarrow{\mathbf{H}}_v\cdot \overset\leftarrow{v}_{j(t+1)} + \overset\leftarrow{b}_v)
\end{split}
\label{equ:bi2}
\end{equation}
The summary of user sequence, i.e., $\overset\rightarrow{u}_{iT_i}$
$\overset\leftarrow{u}_{i1}$, are taken as input for modeling the attention score $a_{jt}$, so as to condition the learning of $a_{jt}$ on the paired user sequence  $\mathbf{x}_i$.

By modeling the attention of each time step in both the user- and item-side networks, our method can capture the interacting dependency and the joint effects of user and item dynamics on user preferences. It thus enable us to gain ``second order'' insights such as how user preferences are determined by the dynamics of user inclinations and the change of item perception/popularity together.

\subsection{Feature Encoder}
\label{sec:HFE}
We now introduce Feature Encoder, which is used to extract semantically rich information from the input data for learning high-quality hidden-states. 
Here we focus on Feature Encoder for processing item-side input, as features of items are in generally richer than users (e.g., the datasets we will take for validation in section~\ref{sec:experiment}). It is however non-trivial to adapt our method for processing the user-side input when auxiliary features of users are given. 

We consider two structures of feature organizations, namely, flat structure and hierarchy. Formally, let $\mathcal{F}$ denote the set of features organized in a flat structure or a hierarchy. Each item $j\in \mathcal{V}$ is affiliated with a subset of features $\mathcal{F}(j) = \{f_j^1, f_j^2, \ldots, f_j^L\}$. The effect of feature $f_j^k$ is modeled as a linear transformation function, denoted by
$\mathbf{M}_j^k$, that projects the input $\mathbf{x}_{jt}$ for all $1\leq t\leq T_j$ to a new space determined by the feature (i.e., the column space of $\mathbf{M}_j^k$)
\begin{equation}
\mathbf{M}_j^k\cdot \mathbf{x}_{jt}
\end{equation}
The problem is how to combine the effects of different features of $\mathcal{F}(j)$ to project the input for best learning the hidden-states. Considering feature organizations, we design our Feature Encoder as follows.

\smallskip
\noindent\textbf{Flat Feature Encoder.} In the case when features are organized in a flat structure, we simply add the effects of different features together. Formally, for the input $\mathbf{x}_{jt}$, the combined effects of all affiliated features $\mathcal{F}(j)$ are given by 
\begin{equation}
E_v(\mathbf{x}_{jt}) =  \sum\nolimits_{k=1}^L \mathbf{M}_j^k \cdot \mathbf{x}_{jt}
\end{equation}

\smallskip
\noindent\textbf{Hierarchical Feature Encoder.} In the case when $\mathcal{F}(j)$ is a feature hierarchy, let $f_j^1$ be the feature in the leaf layer and $f_j^L$ be the root feature. Intuitively, features in top-layers (close to root in the hierarchy) provide more general description of the item, while those in bottom-layers (close to the leaf layer in the hierarchy) provide more refined description. 
%
%
Inspired by the recursive nature of a hierarchy, we consider the recursive parent-children relationships between features in connected layers from the root to leaf layer. In every two connected-layers, the input will be first projected by the parent feature, then by the child feature. By doing so, they will be first mapped to a more general feature space, and then mapped to a more semantically refined feature space. The effects of all affiliated features in different layers will be combined recursively, such that the input can be sequentially mapped to more refined spaces. 

Formally, for the input $\mathbf{x}_{jt}$, the combined effects of all affiliated features $\mathcal{F}(j)$ are given by 
\begin{equation}
E_v(\mathbf{x}_{jt}) = (\mathbf{M}_j^1 \cdot (\mathbf{M}_j^2 \ldots \cdot (\mathbf{M}_j^L  \cdot  \mathbf{x}_{jt})\ldots )) = \prod\nolimits_{k=1}^L \mathbf{M}_j^k \cdot \mathbf{x}_{jt}
\end{equation}

\subsection{End-to-End Parameter Learning}
Given the training data $\mathcal{D}_{train}$ containing $N$ instances in the form of $(i, j, r_{ij}, time stamp)$, IARN learns the involved parameters by minimizing the mean squared error loss function:
\begin{equation}
\mathcal{J} = \frac{1}{N}\sum_{r_{ij}\in \mathcal{D}_{train}} (\tilde{r}_{ij} - r_{ij})^2
\label{equ:objective}
\end{equation}
Since all the modules and the above loss function are analytically differentiable, IARN can be readily trained in an end-to-end manner. In the learning process, parameters are updated by the back-propagation through time (BPTT) algorithm \cite{werbos1988generalization} in the recurrent layers of the Attention-gated Recurrent Module and the Interacting Attention Module, and by normal back-propagation in other parts. We use RMSprop \cite{tieleman2012lecture} to adaptively update the learning rate, which has proven to be highly effective for training neural networks.  To prevent over-fitting, we use dropout \cite{srivastava2014dropout} to randomly drop hidden units of the network in each iteration during the training process. 


\subsection{Comparison with Recurrent Network based Methods}

\noindent\textbf{Comparison with RNN- and LSTM-backbone.} One could argue that our framework can also employ two RNN or LSTM as the backbone for user- and item-side recurrent networks. However, the major downside of RNN- and LSTM-backbone is two-fold. First, RNN- and LSTM-backbone cannot provide interpretable recommendation results either due to the lack of gates (RNN-backbone), or the gates modeled as multi-dimensional vectors (LSTM-backbone). In contrast, gates in IARN are represented by attention scores in scalar values, therefore IARN can provide meaningful interpretations on the relevance of each time step for recommendation. Second, RNN- or LSTM-backbone models user dynamics and item dynamics separately, thus can only learn fixed attention scores for each user and item. The attention scores for a specific user (item) actually indicate the general importance (e.g., the frequency) of each item (user) in purchased history of this user (item), which may not be effective in predicting specific user-item interactions. Unlike them, the novel attention scheme designed for IARN can learn different attention scores for an individual user (item) when interacting different items (users), thus can model the dependencies between user and item dynamics. In addition to the above, when compared with LSTM-backbone, IARN has less parameters, so is less prone to be over-fitting. Moreover, IARN uses the bi-directional recurrent network to model attention gates, which helps to maximize the utility of the input data.  


\smallskip\noindent\textbf{Comparison with TAGM.} IARN is inspired by the Temporal Attention Gated Model (TAGM) \cite{pei2017temporal} recently proposed for sequence classification. IARN inherits the bi-directional attention gates from TAGM, however, our attention scheme is specifically designed with the purpose of recommendation in mind. The nature of recommendation requires proper modeling user-item interactions, for which we design the Interacting Attention Module for modeling the interacting attention for both users and items. This allows IARN to capture the dependencies between user and item dynamics, making IARN particularly suitable for modeling user-item interactions.

\section{Experiments and Results}
\label{sec:experiment}
In this section, we conduct experiments to evaluate the  performance of IARN on six real-world datasets. We aim to answer the following research questions: (1) How do the interacting attention scheme and feature encoder of IARN contribute to recommendation performance and interpretability? (2) How effective is IARN compared to state-of-the-art recommendation methods in both rating prediction and personalized ranking? They will be addressed by section~\ref{sec:result1} and section~\ref{sec:result2}, respectively. 

\subsection{Experimental Setup}

\smallskip\noindent\textbf{Datasets.} To evaluate the effectiveness of IARN, we utilize six real-world datasets, namely Netflix prize dataset, MovieLens, and four Amazon Web store datasets introduced by McAuley et al. \cite{mcauley2015image}, i.e., Electronic, Home,  Clothing, Sport. Each data point in these datasets is a tuple -- (user id, item id, rating, time stamp).  Specifically, the Netflix dataset is a large movie rating dataset scaled from 1 to 5 with a step size of 1, which is collected between November 1999 to December 2005. MovieLens is also a personalized movie rating dataset collected from September 1995 to March 2015 with ratings ranging from 0.5 to 5.0 with a step size of 0.5. Besides, it also contains for each movie the genre information as features in a flat structure. The Amazon Web store datasets are collected from Amazon\footnote{https://www.amazon.com/}, which is a large on-line shopping website, including electronics, clothing, etc. The time span is from May 1996 to July 2014. In addition, there is an item category hierarchy associated with each of the four datasets. We sample the datasets such that only users and items with more than 3 ratings are preserved. Table \ref{tab:statistics} summarizes the statistics of all the considered datasets. 
\begin{table}
\centering
\addtolength{\tabcolsep}{-0.9mm} 
\caption{The statistics of datasets, where \#U\_av\_T (\#I\_av\_T) is the average length of sequences w.r.t. users (items);   }\label{tab:statistics}
\vspace{-0.15in}
\begin{tabular}{l|rrrrrr}
\toprule
\textbf{Datasets} &\#User &\#Item &\#Rating &\#Feature &\#U\_av\_T &\#I\_av\_T  \\\midrule
Netflix  &17,043 &9,598 &721,017 &-- &36.45 &64.73\\
MovieLens &9,737 &5,121 &316,891 &19 &31.07 &59.08 \\
Electronic &11,117 &15,985 &136,998 &590 &11.35 &7.89\\
Home &15,745 &19,383 &201,660 &883 &11.62 &9.44\\
Clothing &19,939 &20,785 &135,128 &690 &6.07 &5.82 \\
Sport &11,723 &13,811 &127,178 &1,130 &9.82 &8.33 \\
\bottomrule
\end{tabular}
\vspace{-0.1in}
\end{table}

\smallskip\noindent\textbf{Comparison Methods.} We compare with the following state-of-the-art algorithms, 1) \textbf{MF} \cite{koren2009matrix}: matrix factorization as the basic latent factor model (LFM) aiming at rating prediction; 2) \textbf{BPR} \cite{rendle2009bpr}: Bayesian personalized ranking as the basic LFM designed for item ranking; 
3) \textbf{TimeSVD++} \cite{koren2009collaborative}: LFM  with the incorporation of temporal context; 4) \textbf{HieVH} \cite{sun2017exploiting}: LFM integrating feature hierarchies; 5) \textbf{Item2Vec} \cite{barkan2016item2vec}: the basic neural network (NN) model; 6) \textbf{NCF} \cite{heneural}: neural collaborative filtering replacing the inner product with non-linear network layers for item ranking; 7) \textbf{MP2Vec} \cite{vasile2016meta}: NN model considering auxiliary features. Note that methods designed for incorporating feature hierarchies can also handle features in a flat structure, by considering all features in the same level; similarly, methods designed for incorporating features in a flat structure can also handle feature hierarchies by flattening them into flat structures, with the loss of certain structural information. 

To investigate the effect of attention-gates and our novel attention scheme, we also compare the following IARN variants using different recurrent networks as the backbone, a) \textbf{RNN-backbone}: the basic variant using RNN as the backbone of user- and item-side recurrent neural networks; b) \textbf{LSTM-backbone}: the variant using LSTM as the backbone; 
c) \textbf{TAGM-backbone}: the variant using TAGM as the backbone; d) \textbf{IARN-Plain}: the variant of our proposed attention-gated recurrent networks integrated with the interacting attention scheme; 
e) \textbf{IARN}: the upgraded version of IARN-Plain by fusing auxiliary features.  Note that LSTM-backbone is similar to  \cite{wu10recurrent} which also employs LSTM as the backbone; while TAGM-backbone is a completely new method  which is adapted from TAGM for recommendation. 
Given their same goal in modeling temporal dynamics, we compare them together. 


\smallskip
\noindent\textbf{Evaluation Metrics.}
We adopt Root Mean Square Error (RMSE) and Area Under the ROC Curve (AUC) to measure the performance of rating prediction and personalized ranking, respectively. 
The smaller RMSE and the larger AUC, the better the  performance. We split all the datasets into training and test data according to the following time stamps: June 1st, 2005 for Netflix dataset; January 1st, 2010 for MovieLens dataset; and January 1st, 2014 for the four Amazon datasets. The data before these time stamps are treated as training data, while the rest are considered as the test data. 

\smallskip\noindent\textbf{Parameter Settings.} We empirically find out the optimal parameter settings for each comparison method. For all the methods, we set the dimension of the latent factor $d = 25$ on Netflix and MovieLens datasets, and $d = 50$ on the four Amazon datasets.  
We apply a grid search in $\{10^{-5}, 10^{-4}, 10^{-3}, 10^{-2}, 10^{-1}\}$ for the learning rate and regularization coefficient. 
For TimeSVD++, $decay\_rate = 0.4; bin = 30$. For HieVH, $\alpha = 0.01$. For MP2Vec, $\alpha = 0.1$.  
For all recurrent networks mentioned in this work (RNN-backbone, LSTM-backbone, TAGM-backbone, IARN) as well as NCF, the number of hidden units is set to 64 which is selected as the best configuration from the option set \{32, 64, 128\} based on a held-out validation set. To avoid  potential over-fitting, the dropout value is validated from the option set \{0.00, 0.25, 0.50\}. Model training  is performed using a RMSprop stochastic gradient descent optimization algorithm with mini-batches of 50 pairs of user-item interactions. All the gradients are clipped between -10 and 10 to prevent exploding~\cite{gradientclip}. 

\subsection{Effects of Attention and Feature Encoder}
\label{sec:result1}

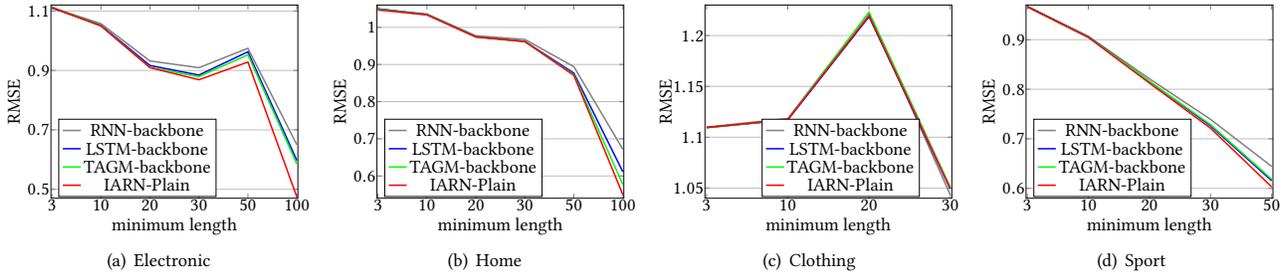
\begin{figure*}[!t]
	\centering
    \subfigure[Electronic]{
		\begin{tikzpicture}[scale=0.4]
		\begin{axis}[
		xlabel=minimum length,
		ylabel=RMSE,
		xmin=0.98,
		xmax=6.02,
		ymin=0.47,
		ymax=1.12,
		xtick={1, 2, 3, 4, 5, 6},
		xticklabels={3,10,20,30,50,100},
		ytick={0.5, 0.7,0.9,1.1},
		legend style={at={(0.35,0.41)},anchor= north, font=\Huge, legend columns= 1},
		yticklabel style = {font=\Huge,xshift=0.5ex},
		xticklabel style = {font=\Huge,yshift=0.5ex},
		xlabel style = {font= \Huge},
		ylabel style = {font= \Huge},
		width=0.55\textwidth,
		height=.45\textwidth,
        ymajorgrids=true
		]
		\addplot[color=gray, line width = 0.5mm] plot coordinates {
			(1, 1.1118)
			(2, 1.0572)
			(3, 0.9319)
			(4, 0.9092)
			(5, 0.9745)
			(6, 0.6499)
		};
		\addplot[color=blue, line width = 0.5mm] plot  coordinates {
			(1, 1.1101)
			(2, 1.0515)
			(3, 0.9166)
			(4, 0.8844)
			(5, 0.9627)
			(6, 0.5967)
		};
        \addplot[color=green, line width = 0.5mm] plot  coordinates {
			(1, 1.1130)
			(2, 1.0502)
			(3, 0.9099)
			(4, 0.8791)
			(5, 0.9526)
			(6, 0.5849)
		};
		\addplot[color=red, line width = 0.5mm] plot coordinates {
			(1, 1.1117)
			(2, 1.0495)
			(3, 0.9095)
			(4, 0.8687)
			(5, 0.9282)
			(6, 0.4739)
		};
		\legend{RNN-backbone\\LSTM-backbone\\TAGM-backbone\\IARN-Plain\\}
		\end{axis}
		\end{tikzpicture}
	}
    \subfigure[Home]{
		\begin{tikzpicture}[scale=0.4]
		\begin{axis}[
		xlabel=minimum length,
		ylabel=RMSE,
		xmin=0.98,
		xmax=6.02,
		ymin=0.54,
		ymax=1.06,
		xtick={1, 2, 3, 4, 5, 6},
		xticklabels={3,10,20,30,50,100},
		ytick={0.6,0.7,0.8,0.9,1.0},
		legend style={at={(0.35,0.41)},anchor= north, font=\Huge, legend columns= 1},
		yticklabel style = {font=\Huge,xshift=0.5ex},
		xticklabel style = {font=\Huge,yshift=0.5ex},
		xlabel style = {font= \Huge},
		ylabel style = {font= \Huge},
		width=0.55\textwidth,
		height=.45\textwidth,
        ymajorgrids=true
		]
		\addplot[color=gray, line width = 0.5mm] plot coordinates {
			(1, 1.0492)
			(2, 1.0362)
			(3, 0.9772)
			(4, 0.9674)
			(5, 0.8941)
			(6, 0.6717)
		};
		\addplot[color=blue, line width = 0.5mm] plot coordinates {
			(1, 1.0499)
			(2, 1.0341)
			(3, 0.9747)
			(4, 0.962)
			(5, 0.8767)
			(6, 0.6111)
		};
        \addplot[color=green, line width = 0.5mm] plot coordinates {
			(1, 1.0485)
			(2, 1.0342)
			(3, 0.9743)
			(4, 0.9610)
			(5, 0.8726)
			(6, 0.5767)
		};
		\addplot[color=red, line width = 0.5mm] plot coordinates {
			(1, 1.0469)
			(2, 1.0335)
			(3, 0.9737)
			(4, 0.9619)
			(5, 0.8706)
			(6, 0.5494)
		};
		\legend{RNN-backbone\\LSTM-backbone\\TAGM-backbone\\IARN-Plain\\}
		\end{axis}
		\end{tikzpicture}
	}
    \subfigure[Clothing]{
		\begin{tikzpicture}[scale=0.4]
		\begin{axis}[
		xlabel=minimum length,
		ylabel=RMSE,
		xmin=0.98,
		xmax=4.02,
		ymin=1.04,
		ymax=1.23,
		xtick={1, 2, 3, 4},
		xticklabels={3,10,20,30},
		ytick={1.05, 1.10, 1.15, 1.20},
		legend style={at={(0.55,0.41)},anchor= north, font=\Huge, legend columns= 1},
		yticklabel style = {font=\Huge,xshift=0.5ex},
		xticklabel style = {font=\Huge,yshift=0.5ex},
		xlabel style = {font= \Huge},
		ylabel style = {font= \Huge,yshift=0.5ex},
		width=0.55\textwidth,
		height=.45\textwidth, 
        ymajorgrids=true
		]
		\addplot[color=gray, line width = 0.5mm] plot coordinates {
			(1, 1.1100)
			(2, 1.1184)
			(3, 1.2224)
			(4, 1.0435)
		};
		\addplot[color=blue, line width = 0.5mm] plot coordinates {
			(1, 1.1096)
			(2, 1.1170)
			(3, 1.2186)
			(4, 1.0492)
		};
        \addplot[color=green, line width = 0.5mm] plot coordinates {
			(1, 1.1093)
			(2, 1.1173)
			(3, 1.2230)
			(4, 1.0510)
		};
        \addplot[color=red, line width = 0.5mm] plot coordinates {
			(1, 1.1094)
			(2, 1.1174)
			(3, 1.2197)
			(4, 1.0498)
		};
		\legend{RNN-backbone\\LSTM-backbone\\TAGM-backbone\\IARN-Plain\\}
		\end{axis}
		\end{tikzpicture}
	}
    \subfigure[Sport]{
		\begin{tikzpicture}[scale=0.4]
		\begin{axis}[
		xlabel=minimum length,
		ylabel=RMSE,
		xmin=0.98,
		xmax=5.02,
		ymin=0.58,
		ymax=0.97,
		xtick={1, 2, 3, 4, 5},
		xticklabels={3,10,20,30,50},
		ytick={0.6,0.7,0.8,0.9},
		legend style={at={(0.35,0.41)},anchor= north, font=\Huge, legend columns= 1},
		yticklabel style = {font=\Huge,xshift=0.5ex},
		xticklabel style = {font=\Huge,yshift=0.5ex},
		xlabel style = {font= \Huge},
		ylabel style = {font= \Huge},
		width=0.55\textwidth,
		height=.45\textwidth,
        ymajorgrids=true
		]
		\addplot[color=gray, line width = 0.5mm] plot coordinates {
			(1, 0.9675)
			(2, 0.9068)
			(3, 0.8211)
			(4, 0.7392)
			(5, 0.6441)
		};
		\addplot[color=blue, line width = 0.5mm] plot coordinates {
			(1, 0.9671)
			(2, 0.9058)
			(3, 0.8155)
			(4, 0.7262)
			(5, 0.6155)
		};
        \addplot[color=green, line width = 0.5mm] plot coordinates {
			(1, 0.9664)
			(2, 0.9053)
			(3, 0.8158)
			(4, 0.7285)
			(5, 0.6183)
		};
		\addplot[color=red, line width = 0.5mm] plot coordinates {
			(1, 0.9663)
			(2, 0.9045)
			(3, 0.8124)
			(4, 0.7213)
			(5, 0.6028)
		};
		\legend{RNN-backbone\\LSTM-backbone\\TAGM-backbone\\IARN-Plain\\}
		\end{axis}
		\end{tikzpicture}
	}
	\vspace{-0.15in}
	\caption{Performance (measured by RMSE) of IARN variants with different recurrent networks as the backbone on different configurations of the real-world datasets with varying minimum sequence lengths.} \label{fig:minlength}
\end{figure*}

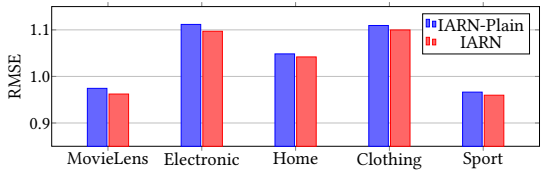
\begin{figure}[!t]
    \centering
     \pgfplotstableread[row sep=\\,col sep=&]{
        cases & without & with \\
        MovieLens     & 0.9744  & 0.9622   \\
        Electronic   & 1.1117 & 1.0970\\
        Home    & 1.0484 & 1.0419   \\
        Clothing  & 1.1094  & 1.0998   \\
        Sport    & 0.9663 & 0.9597 \\
     }\mydata

     \begin{tikzpicture}[scale=0.4]
     \begin{axis}[
     ybar,
     bar width=.65cm,
     width=1.0\textwidth,
     height=.35\textwidth,
     legend style={at={(0.87,0.95)},
        anchor=north,legend columns= 1, font = \Huge},
     symbolic x coords={MovieLens, Electronic, Home, Clothing, Sport},
     xtick=data,
     enlarge x limits=0.15,
     ymin=0.85,ymax=1.15,
     ylabel={RMSE},
     ytick = {0.9,1.0,1.1},
     yticklabels = {0.9, 1.0, 1.1},
     yticklabel style = {font=\Huge,xshift=0.5ex},
     xticklabel style = {font=\Huge,yshift=0.5ex},
     ylabel style ={font = \Huge},
     xlabel style = {font = \Huge},
     ymajorgrids=true
     ]
     \addplot[draw=blue,fill=blue!60!white, thick] table[x=cases,y=without]{\mydata};
     \addplot[draw=red,fill=red!60!white] table[x=cases,y=with]{\mydata};
     \legend{IARN-Plain, IARN}
     \end{axis}
     \end{tikzpicture}
     \vspace{-0.05in}
    \caption{Performance of rating prediction of IARN variants with and without feature encoder on the five datasets.} \label{fig:feature}
    \vspace{-0.15in}
\end{figure}

\noindent\textbf{Attention.} 
In order to investigate the impact of the proposed attention scheme, we compare the performance (measured by RMSE) of IARN-Plain with different recurrent networks as the backbone, including RNN-backbone, LSTM-backbone, and TAGM-backbone. To understand their capability in modeling temporal dynamics of users and item history in different lengths, we test their performance on different configurations of the datasets by constraining the minimum length of user and item input sequences. A grid search in $\{3, 10, 20, 30, 50, 100\}$ is applied for the minimum length of sequences on all the datasets, excluding Clothing and Sport since there are few users possessing long length sequences in these two datasets. Due to space limitation, we only show the results on four Amazon datasets as depicted by Figure \ref{fig:minlength}, however similar observations as below can be obtained on all the datasets. 

As the minimum length of input sequences increases, the performance of all methods generally improves, indicating that sufficient temporal context could ensure recurrent network based methods to better model the dynamics of user preferences.  
The performance of gated recurrent networks, i.e., LSTM-backbone, TAGM-backbone, and IARN-plain, is generally better than the non-gated recurrent network, i.e., RNN-backbone. Such a difference is minor when the minimum sequence length is less than a threshold (e.g., 30), and becomes significant with the further growth of the sequence length. This shows the benefit of gating mechanism in effectively preserving historical information deep into the past for recommendation. The observation further explains the close performance of different methods on Clothing dataset, whose sequences are mostly less than 30 and the average sequence length (i.e., around 6, Table~\ref{tab:statistics}) is significantly smaller than all the other datasets. 

The overall performance of TAGM-backbone and IARN-Plain, is better than that of LSTM-backbone. LSTM-backbone adopts multi-dimensional gates w.r.t. each hidden unit, which can be more easily over-fitted than the (bi-directional) attention-gates employed by TAGM-backbone and IARN-Plain. With respect to attention-gated recurrent networks, IARN-Plain outperforms TAGM-backbone across all different configurations of minimum sequence length. This is mainly due to the fact that TAGM-backbone learns user and item dynamics separately, i.e., only fixed attention scores for user history and item history are learned (LSTM-backbone suffers from the same issue). Whereas equipped with our novel attention scheme, IARN-Plain can adaptively learn different attention scores for user (item) history when the user (item) interacts with different items (users). Such a comparison clearly shows the advantage of our proposed attention scheme for modeling the dependencies between user and item dynamics. 


Overall, IARN-Plain achieves the best performance across all different configurations of all the datasets, especially when the sequence length gets larger. On average, the relative improvements w.r.t. the second best method are 2.54\% with minimum length $= 50$ and 11.65\% with minimum length $= 100$.  This implies the remarkable advantage of IARN-Plain in dealing with long sequences.

\smallskip\noindent\textbf{Feature Encoder.}
We further examine the effectiveness of auxiliary features which are organized in either a flat or hierarchical structure on all the datasets, excluding Netflix which does not contain any auxiliary features. The results are given by Figure \ref{fig:feature}. By integrating auxiliary features IARN outperforms IARN-Plain across all the datasets, with 1.19\% lift ($p$-value $<0.01$) in RMSE on average. This clearly indicates the benefit of considering feature encoder in our proposed IARN approach. 

%

\begin{figure*}
\begin{tikzpicture}
\matrix (a)[row sep=0mm, column sep=0mm, inner sep=1mm,  matrix of nodes] at (0,0) {
	\subfigure[MovieLens]{
    \includegraphics[width=0.48\textwidth]{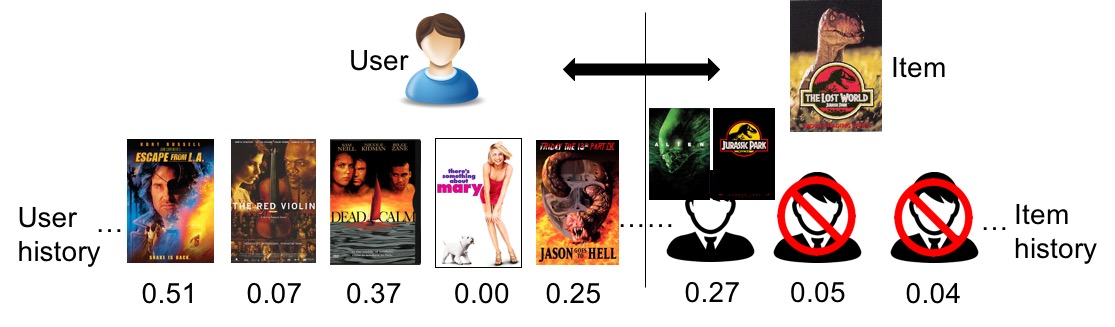}
    } &
    \subfigure[Electronic]{
    \includegraphics[width=0.48\textwidth]{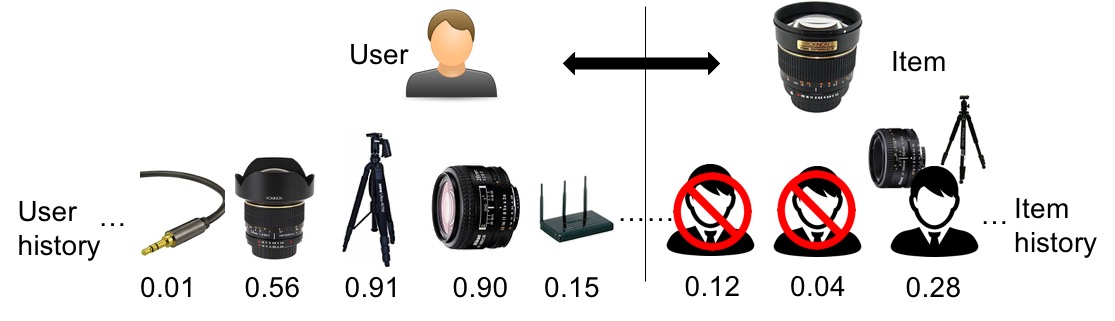}
    } \\
    \subfigure[Home]{
    \includegraphics[width=0.48\textwidth]{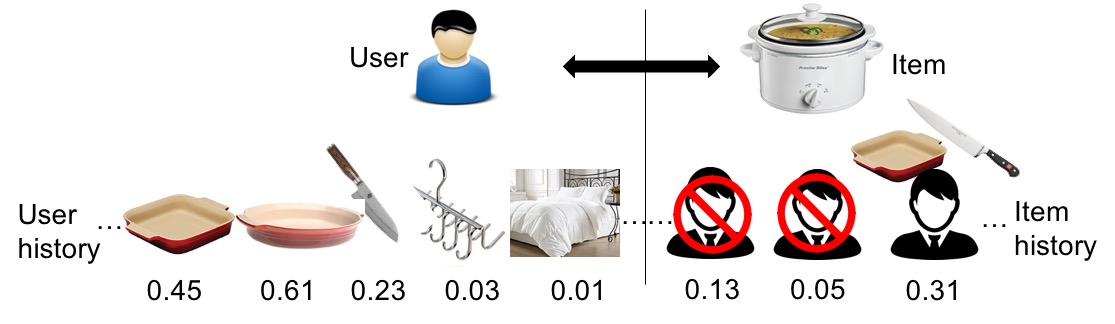}
    } &
    \subfigure[Sport]{
    \includegraphics[width=0.48\textwidth]{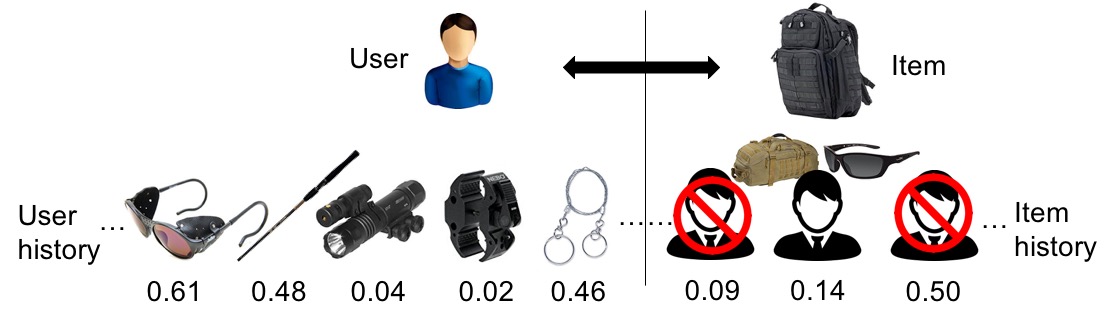}
    } \\\\
};
\draw[thick,densely dashed,gray] (a-1-1.north east) -- (a-2-1.south east);
\draw[thick,densely dashed,gray] (a-2-1.north west) -- (a-1-2.south east);
\end{tikzpicture}
\vspace{-0.2in}
\caption{Examples of attention scores learned by IARN. The target user and item in each sub-figure have an observed interaction. The attention scores of individual time steps are shown under user and item history.}\label{fig:interpretation}
\end{figure*}

\smallskip\noindent\textbf{Interpretation by IARN.}
The attention scores learned by IARN for individual time steps in user and item history can help quantify the relevance of each time steps in user and item history for recommendation. We now qualitatively analyze such attention scores to investigate their effects on providing meaningful interpretations for recommendation. Figure~\ref{fig:interpretation} shows the attention scores learned by IARN on examples of four datasets. 

In each of the four examples, we can observe varying attention scores assigned to different time steps in both user and item history. Such attention scores can effectively capture the target user's preference related to the inherent property of the target item, as inferred from the data. For example in MovieLens dataset, IARN learns high attention scores for the time steps in user history when the user was watching movies of genre ``Adventure'' and ``Action''. These time steps are highly indicative of his potential preference over the target item, i.e., ``The Lost World: Jurassic Park''. In contrast, low attention scores are assigned to those time steps when he was watching movies of other genres, e.g., ``Drama''. IARN thus can selectively memorize most relevant time steps of the user's history in predicting his preference over the target item. Similarly, IARN can also select the most relevant time steps in item history to characterize the inherent genre of the item, i.e., those time steps when it was being watched by users who share the same interest as the target user, i.e., ``Adventure'' and ``Action'' movies (e.g., ``Aliens''). Similar observations can be noted in the other three examples. For instance in the Sport dataset, IARN can infer the most relevant time steps in the user history when the user bought hiking related item; and in the item history when the item was bought by users who like hiking. Such dependency between the relevance of time steps in user history and in item history is highly useful for discovering the link between the target user and item, and thus provides strong interpretations for the recommendation results. 

%
\begin{table*}
\centering
\addtolength{\tabcolsep}{-0.6mm}
\caption{Performance of rating prediction (measured by RMSE) and personalized ranking (measured by AUC) of all comparison methods on the six real-world datasets. The best performance is boldfaced; the runner up is labeled with ``*''. The results of HieVH and MP2Vec on Netflix is not available (marked by ``--'') due to the lack of feature information in the Netflix dataset. }\label{tab:comparison}
\vspace{-0.15in}
\resizebox{1.0\linewidth}{!}{
\begin{tabular}{l|cccc|cccccccc}
\toprule
\multirow{2}{*}{\textbf{Datasets}} & \multicolumn{4}{c}{Rating prediction: RMSE} & \multicolumn{8}{c}{Personalized ranking: AUC}  \\
& \textbf{MF} & \textbf{TimeSVD++} & \textbf{HieVH}  & \textbf{IARN} & \textbf{MF} & \textbf{BPR} & \textbf{TimeSVD++} & \textbf{HieVH} & \textbf{Item2Vec}  & \textbf{NCF} & \textbf{MP2Vec} & \textbf{IARN} \\
\midrule
Netflix &1.1828   &1.1243* &--  &\textbf{1.0582}  &0.6147 &0.7414 &0.6289 &-- &0.7642 & 0.7654* & -- & \textbf{0.7901}  \\
MovieLens &1.1319   &1.0623 &1.0326*  & \textbf{0.9622}  &0.6305 &0.6971 &0.6504 &\textbf{0.7214} &0.7130 &0.7128 & 0.7135* & 0.7135* \\
Electronic  &1.3213 &1.3152 &1.1694*  &\textbf{1.0970} &0.5757 &0.6699 &0.5820 &0.7257* & 0.6794 & 0.7052 & 0.7072& \textbf{0.7359} \\
Home &1.2010   &1.1974 &1.1231*  & \textbf{1.0419} &0.5305 &0.6341 &0.5520 &0.7132* &0.6902 & 0.6973 & 0.7007 & \textbf{0.7210}\\
Clothing &1.3587   &1.2851 &1.2283* & \textbf{1.0998}&0.5092 &0.6246 &0.5205 &\textbf{0.7011} &0.6717 & 0.6720 & 0.6919&{0.7004}* \\
Sport  &1.2021   &1.1690 &1.1279*  & \textbf{0.9597} &0.5489  &0.6492 &0.5515 &0.6962* &0.6787 & 0.6759& 0.6792 &\textbf{0.6975} \\
\bottomrule
\end{tabular}}
\end{table*}

\subsection{Comparative Results}
\label{sec:result2}

\noindent\textbf{Rating Prediction.}
The left side of Table \ref{tab:comparison} presents the rating prediction performance on the six real-world datasets. BPR, Item2Vec, NCF, and MP2Vec are excluded since RMSE cannot be applied to these methods. BPR and NCF optimize ranking based objective function. Item2Vec and MP2Vec learn the embeddings of items and then adopt the similarity score between item embeddings to predict recommendations, instead of minimizing the difference between the real ratings and the estimated ratings. 
Several interesting observations can be obtained. 

It is unsurprising that MF -- as the basic LFM --  considering no auxiliary information, performs the worst among all the considered methods. 
By integrating temporal context into LFM, TimeSVD++ outperforms MF. This confirms that modeling temporal dynamics of user preferences can significantly improve the recommendation performance. 
HieVH is also a LFM based approach, which takes into account the influence of both vertical and horizontal dimensions of feature hierarchies on recommendation.   It outperforms MF, and even slightly exceeds TimeSVD++, confirming the effectiveness of auxiliary features for better recommendation.

Our proposed approach -- IARN, consistently outperforms the other methods in the comparison pool, with an average performance gain (w.r.t. the second best method) of 8.58\% on RMSE. Pair-wised t-test demonstrates that the improvements of IARN on all the datasets are significant ($p-$value$<0.01$). Such big enhancement clearly shows the effectiveness of the integration of interacting attention scores as well as auxiliary features. 


\smallskip\noindent\textbf{Ranking Performance.}
We further evaluate the ranking quality of items recommended by the methods in the comparison pool. Results are shown on the right side of Table~\ref{tab:comparison}.
A number of meaningful findings can be noted from the table. 

In terms of the LFM based methods, TimeSVD++ and HieVH outperform MF by taking temporal context and feature hierarchies into account, respectively. This observation further verifies 
 the usefulness of the two types of side information for better recommendations. For NN based method, the fact that the performance of MP2Vec is better than that of Item2Vec and NCF also helps to reach the same conclusion, as MP2Vec considers auxiliary features while Item2Vec and NCF do not. The superior performance of NCF over Item2Vec shows the effectiveness of hidden layers in neural networks for modeling non-linear user-item interactions. In both LFM based methods and NN based methods, those specifically designed for personalized ranking, i.e., BPR and NCF, perform better than methods for rating prediction, i.e., MF and Item2Vec, 
%
which strongly confirms the conclusion that methods designed for personalized ranking are more efficient than rating prediction methods for the item ranking problem \cite{rendle2009bpr}.  

Our proposed approach IARN generally achieves the best performance on item ranking when compared with the other considered methods. This demonstrates the effectiveness of IARN in modeling user and item dynamics for improving recommendation performance. However, the performance improvements of IARN on ranking prediction is far behind those on rating prediction. The underlying explanation is that the objective function of IARN aims to minimize the squared error between the observed ratings and the estimated ratings, which is just in accordance with the definition of RMSE. IARN is therefore more effective on rating prediction. We leave it as future work the improvement of IARN on item ranking.

\section{Conclusions}

User preferences often evolve over time, 
thus modeling their temporal dynamics is essential for recommendation.  
This paper proposes the Interacting Attention-gated Recurrent Network (IARN) to accommodate temporal context for better recommendation. IARN can not only accurately measure the relevance of individual time steps of user and item history for recommendation, but also capture the dependencies between user and item dynamics in shaping user-item interactions. We further show that IARN can be easily integrated with auxiliary features for enhanced recommendation performance. 
Extensive validation on six real-world datasets demonstrates the superiority of IARN against other state-of-the-art methods. 
For future work, we intend to further improve the effectiveness of IARN on the item ranking problem. 

\section*{Acknowledgement}
This work is partially funded by the Social Urban Data Lab (SUDL) of the Amsterdam Institute for Advanced Metropolitan Solutions (AMS).
This work is partially supported by the SIMTech-NTU Joint Lab on Complex Systems.
\bibliographystyle{ACM-Reference-Format}
\bibliography{sigproc} 

\end{document}